# New Synthesis Method for the Growth of Epitaxial Graphene

X. Z. Yu[1,2], C. G. Hwang[1], C. M. Jozwiak[1], A. Köhl[1,*], A. K. Schmid[3] and A. Lanzara[1,4,a]

[1]*Materials Sciences Division, Lawrence Berkeley National Laboratory, Berkeley, California 94720, USA*

[2]*Laboratory of Condensed Matter Spectroscopy and Opto-Electronic Physics, Department of Physics, Shanghai Jiao Tong University, 1954 Hua Shan Road, Shanghai 200030, China*

[3]*National Center for Electron Microscopy, Lawrence Berkeley National Laboratory, Berkeley, California 94709, USA*

[4]*Department of Physics, University of California, Berkeley, California 94720, USA*

## Abstract

As a viable candidate for an all-carbon post-CMOS electronics revolution, epitaxial graphene has attracted significant attention. To realize its application potential, reliable methods for fabricating large-area single-crystalline graphene domains are required. A new way to synthesize high quality epitaxial graphene, namely "*face-to-face*" method, has been reported in this paper. The structure and morphologies of the samples are characterized by low-energy electron diffraction, atomic force microscopy, angle-resolved photoemission spectroscopy and Raman spectroscopy. The grown samples show better quality and larger length scales than samples grown through conventional thermal desorption. Moreover the graphene thickness can be easily controlled by changing annealing temperature.

[a] Corresponding author, E-mail: ALanzara@lbl.gov

[*] Currently at: Department of Physics, University of Wurzburg, Wurzburg 97070, Germany

## Introduction

Graphene, a two-dimensional array of carbon atoms in honeycomb lattice, has been theoretically studied for decades in terms of the fundamental building block of carbon based materials, such as graphite and carbon nanotubes [1, 2]. However, the first observation of freestanding graphene was not realized until 2004 with mechanical exfoliation method by peeling graphene flakes from a bulk graphite crystal onto $SiO_2$ substrate [3]. Since then, a variety of novel properties, including quantum Hall effects, and relativistic quasiparticles with a group velocity of 1/300c (where c is the speed of light) [4-6] have been observed. These unique properties, together with the high values of conductance, mobility, and mechanical strength [4, 7-8] make graphene a promising material for a wide variety of new technological applications [9-11], such as post-CMOS digital electronics, single-molecule gas sensors, spintronic devices, etc.

To realize the application potential of graphene, reliable methods for fabricating large-area single-crystalline graphene domains are required. The most promising approach in this respect seems to be the controlled graphitization of SiC surfaces [9, 12]. In this method, single layer and/or multilayer graphene can be grown by sublimating Si atoms from SiC substrates at high temperature. In spite of a lot of efforts devoted to the improvement of synthesis methods to form large areas of uniform, electronic grade graphene by the thermal desorption process is still very challenging. In this paper, we will give a short review on the current state of epitaxial graphene research, and we will then introduce a new developed synthesis method, the "*face-to-face*" method, which allows preparing good quality monolayer, bilayer and three layer epitaxial graphene samples on 6H-SiC(0001) substrate.

Although the epitaxial growth of graphene on semiconducting substrates was known as early as 1975 [13], renewed interest in this field over the past few years is certainly due to: the discovery of the fractional quantum Hall effect in freestanding graphene, the observation of 2D electron gas behavior also in the epitaxial samples [9] as well as the appealing possibility to incorporate the existing silicontechnology to mass produce and pattern epitaxial graphene.

As of today, a variety of different methods have been developed to grow graphene epitaxially, both on semiconducting or metallic substrates that have good lattice matching with graphene. All these different methods can be classified into three categories: (a) thermal decomposition of carbide (particularly SiC); (b) cracking of hydrocarbon gas on other carbide substrates (TiC, TaC, WC, etc) and metallic substrates (Ni, Pt); (c) segregation of the carbon atoms from the bulk (Ru(001)).

In this paper we will focus on the epitaxial growth of graphene on SiC substrate, being this most relevant to our newly developed method.

**Epitaxial Graphene on SiC**

Although growing thick graphite samples on SiC has been a well-known process for many years [13, 14, 15], it is not until recently that the thickness was pushed down to few layers and a full characterization of the high quality graphene sample has been carried out [9, 16, 17]. The growth of epitaxial graphene on SiC is based on thermal decomposition of the SiC substrate. Both e-beam heating as well as resistive heating have been used, but no difference seems to arise from the different heating methods [18]. In order to avoid contaminations the heating is usually performed in ultra-high vacuum (UHV) environment. Similar results have been observed for high and/or low base pressure growth but so far no comparative study about the influence of the background pressure in the vacuum chamber has been conducted. From the molar densities one can calculate that approximately three bilayers of SiC are necessary to set free enough carbon atoms for the formation of one graphene layer [18]. The growth of graphene can take place on both the (0001) (silicon-terminated) or (000-1) (carbon-terminated) faces of 4H-SiC and 6H-SiC wafers. The main difference lies in the sample thickness that one can achieve. In the case of silicon face, the growth is slow and terminates after relatively short time at high temperatures giving rise to very thin samples, up to a monolayer. On the contrary, in the case of the carbon face, the growth does not self-terminate giving rise to relatively thick samples (approximately 5 up to 100 layers) [19] with larger orientational and turbostatic disorder [20, 21, 22].

Below we will briefly review the growth conditions for epitaxial graphene on SiC substrate. For a detailed review see [19].

**Epitaxial graphene on SiC(0001) face**

Hydrogen etching is performed as a normal routine of pretreatment to remove scratches from polishing and oxides and leave a surface with highly uniform, atomically flat terraces. Although it is believed that larger graphene sheets should be obtained with a smoother graphitization surface [23], the relevance of pre-graphitization SiC surface to better graphene order has not been substantiated. Recently, the relation between initial surface morphology and sample quality has been discussed by Ohta *et al*. [24]. They observed the formation of graphene on SiC by Si sublimation in Ar atmosphere, and identified two types of monolayer graphene with different shapes. It was noticed that large graphene sheets preferred to grow along the triple bilayer SiC steps, while narrow graphene ribbons formed following the surface of single bilayer SiC height. The dependence between growth mechanisms and initial surface morphology indicates the effects of $H_2$ etching on the formation of graphene. The result suggests that by minimizing the number of single bilayer SiC steps with $H_2$ etching, better graphene sample should be achieved.

To compensate the depletion of Si during the pre-cleaning process, external Si flux is applied before the SiC substrate is heated to higher temperature to grow graphene. A number of surface reconstructions prior to graphitization have been observed and studied by low-energy electron diffraction (LEED) [14, 16]. Fig. 1 shows LEED patterns obtained at different stages during the growth of graphene. The initial Si-rich ($3 \times 3$) phase can be obtained by exposing a well-outgassed SiC surface to a Si flux at a temperature of 800 ℃. A subsequent 5 min annealing at 1000 ℃ in the absence of Si flux gives rise to the sharp pattern shown in panel (a), corresponding to the $1 \times 1$ spots of SiC. Further annealing for 5 min at 1100 ℃ produces the $(\sqrt{3} \times \sqrt{3})R30$ reconstruction shown in panel (b). Finally, the complex $(6\sqrt{3} \times 6\sqrt{3})R30$ pattern shown in panel (c) appears after 10 min annealing at 1250℃, demonstrating the formation of graphene. In different experiments, the annealing temperature can be various. It should be noticed that the first carbon layer grown on the Si-face of SiC, referred as buffer layer, is not graphene.

Although the atomic arrangement of this layer is identical to that of graphene, however, unlike graphene, one third of the carbon atoms of this layer are covalently bonded to the underlying Si atoms of the topmost SiC layer. The π-band is developed between buffer layer and following graphene layers.

A review of current publications shows that the domain size of epitaxial graphene grown in UHV wouldn't be larger than hundred nanometers [18]. The high sublimation rate of Si in vacuum prohibits the carbon atoms from better rearrangement, resulting in the roughening and nucleation of small graphene flakes. Ar-assisted epitaxial growth method was proposed and reported by Emtsev *et al*. [25]. They tested a wide range of annealing temperatures and Ar pressures, and observed significant improvement of surface morphology in high pressure Ar atmosphere. The present of argon molecules hinders the transport of silicon atoms away from the SiC surface, reducing the overall sublimation rate and allowing an increase in graphitization temperature. Arrays of parallel terrace up to 3 um wide and more than 50 um long single layer graphene was produced by this method. The significant improvement of sample morphology is shown in Fig. 2. Moreover, enhanced carried mobility was detected, suggesting better electronic property of the sample. The control of the graphene formation also can be realized by supplying excess Si as reported in Ref. 26. Disilane ($Si_2H_6$) was used as a source of Si to generate a background pressure. Large scale of single layer graphene was produced at 1300℃ in a disilane background pressure of $2 \times 10^{-5}$ Torr.

**Epitaxial graphene on SiC(000-1) face**

Compared to the Si-face, graphene on SiC(000-1) face has been less studied, due to the early conclusion that C face films were of poor quality and rotationally disordered [27]. However, a detailed study indicates that the azimuthal disorder detected by low energy electron diffraction (LEED) is not random [28]. In multiple layer graphene grown on SiC(000-1) face, epitaxial layers can orient either in the 30° phase or in the ±2° phase with respect to the substrate, which is indicated in Fig. 3. The different orientations between adjacent layers cause them to decouple form each other, forming a system which preserves the transport and electronic properties of free-standing monolayer graphene [12, 29]. Due to this unique property, C-face grown graphene attracts more and more attention.

The major problem with the growth of epitaxial graphene on C-face of SiC is the lack of precise control in sample thickness. The controllability has been demonstrated by Camara *et al.* [30] lately. Monolayer graphene of several $\mu m^2$ has been successfully produced by covering the SiC substrate with a graphite cap to lower desorption of Si species. The annealing was performed in a commercial radio frequency heated furnace under vacuum ($\sim 10^{-6}$ Torr).

**The *Face-to-Face* Method**

As indicated in the review above, high quality epitaxial graphene sheets in large scales now can be achieved. However, the growth process gets more complicated because more parameters are introduced into the system, such as annealing atmosphere. In the following part, we will report a new synthesis method, the "*face-to-face*" method, which is straightforward, simple, and economical, and yet yields good quality graphene of large length scales. Here two SiC substrates are placed one on top of the other face to face, with a small gap in between, and are then heated simultaneously. The grown samples show larger terraces sizes and better homogeneity than the one obtained by annealing SiC in ultra high vacuum, as demonstrated by LEED, atomic force microscopy (AFM), angle resolved photoemission spectroscopy (ARPES) and Raman measurements performed on the as grown sample. Moreover we show that the graphene thickness can be easily and well controlled by changing the annealing temperature. The results suggest that the method has potential for efficient production of graphene base devices.

Figure 3(a) shows schematic diagram of our vacuum furnace with a base pressure $\leq 1\times10^{-6}$ Torr, maintained by 30 *l*/s hybrid turbo pump (HTP). Two rectangular pieces of SiC are stacked in pairs with spacers of Ta foil at the edges for a 25 micron gap between the two inner surfaces, the key idea of this method from which the name "*face-to-face*". The pieces are oriented so that the Si-terminated surfaces face each other. The two ends of the stack are wrapped by "L"-shaped pieces of Ta foil at each end and connected to electrodes allowing simultaneous parallel resistive heating of the samples as shown in panel (b) and (c). Temperature was monitored by an infrared optical pyrometer set to an emissivity of 0.96.

The simple geometry of this method results in two important effects. First, at temperatures below 1500 °C (before graphene growth has begun), both pieces act as sources as well as sinks of SiC on the opposing surface (Fig. 4a). Large atomically flat terraces form during this annealing process without careful hydrogen etching which is usually applied in other approaches as a pre-treatment for graphene preparation. By first creating large, flat substrate terraces, the eventual graphene layer can have similarly large terrace sizes. Second, the close proximity of the two surfaces partially traps Si atoms which sublimate from each heated surface, increasing the local partial pressure of Si within the gap. The pressure of Si vapor next to the surfaces of the SiC crystals restricts the net rate of Si sublimation from the substrates (Fig. 4) and allows large pieces of graphene to be formed as described above. Thus, the same effect as in previous methods aiming to improving graphene annealing conditions by reducing Si sublimation (ambient pressures of Ar or disilane) is achieved in a very simple manner. We note that as Si can escape from the gap near the edges of the pieces, control of the Si sublimation from the edge of the substrates is reduced, resulting in thicker and poorer quality of graphene around the edges.

The substrates used in the experiments were cut into 4×6 mm$^2$ using a diamond saw from n-type 6H-SiC (0001) single crystalline wafers[1]. The substrates were degassed at 700 °C for 4 hours followed by annealing at elevated temperatures of 1530 – 1700 °C. During the initial growth stage at 1500 °C, epitaxial SiC layers are nucleated on the two substrates creating atomically flat SiC with large terrace sizes. The sublimation of Si prevails over decomposition of SiC with higher annealing temperature than 1530 °C, which results in the formation of micrometer scale graphene sheets. As in other methods, higher temperatures lead to thicker, or more layers, of graphene. In the present method, 1530 °C results in single layer graphene, and 1700 °C results in triple layer graphene.

**Characterization of _face-to-face_ sample.**

---

[1] Cree, Inc. (Catalog Number: W6NRD0X-0000)

Graphene samples were transferred *ex-situ* to be characterized by several methods. Surface morphology and structure were analyzed by AFM[2] and low-energy electron diffraction (LEED). Thickness and quality of the graphene were characterized with ARPES and Raman spectroscopy. ARPES was conducted in an ultra-high vacuum chamber with base pressure $< 8 \times 10^{-11}$ Torr with He II excitation (hv = 40.8 eV) and hemispherical electron analyzer[3]. Raman spectra were collected using 532 nm laser (2.33 eV) with 1 mW incident power. The laser spot size is around 1 μm in diameter. All measurements were performed at room temperature.

Figure 5(a) – (c) show the structural evolution of samples prepared by the "*face-to-face*" method at three different temperatures measured by AFM. Panel (a) shows the morphology of the sample after degassing procedure to remove any remaining chemical solutions and debris, and to thermally stabilize the sample before annealing it further at higher temperatures for graphene growth. The sample does not show any significant difference from that of untreated SiC substrate. Scratches of various depths still exist and are randomly distributed over the surface. An AFM image of the sample annealed further at 1500 °C is presented in panel (b). We find an atomically flat SiC substrate with micrometer scale terraces separated by meandering step edges, whose height ranges from 2 nm to greater than 5 nm, depending on terrace width. Such modification of the surface is not because of the formation of graphene, but due to the annealing of SiC described above and shown in Fig. 4. This is also confirmed by LEED image, where only a 1 × 1 phase corresponding to bulk SiC is observed (inset of panel (b)). This result can be explained by the epitaxial growth model by thermal equilibrium [31, 32].

When temperature is increased to 1530 °C, the process changes from SiC annealing to the formation of graphene as diagrammed in "stage II" of Fig. 4. At this temperature, the extra C atoms on the surface left from Si sublimation reorganize into the crystal structure. Without the reduction of the Si sublimation rate provided by the current geometry, too much Si would escape at this temperature, resulting in very thick and inhomogeneous graphene. By controlling the Si sublimation rate, precise control of the graphene growth and thickness at relatively high temperature is achieved. Fig. 5(c) displays an AFM image and $6\sqrt{3} \times 6\sqrt{3}$ reconstruction LEED pattern of a single layer of graphene. The terrace edges become less smooth compared with the SiC shown in panel (b) indicating that thermal decomposition of graphene does not necessarily follow the substrate morphology. Other characteristic features of graphene grown using our method are pits and curving steps. Pits have been proposed to be formed when the domains of the carbon rich surface, the precursor of graphene layers grown on SiC, pin the decomposing surface steps [33].

---

[2] Digital Instrument Dimension 3100 Scanning probe microscopy

[3] SPECS Phoibos 150

For comparison, single layer graphene made by the conventional thermal decomposition method as described in detail in Ref. [16] is also characterized by AFM and LEED as shown in Fig. 5(d). The density of pits and the homogeneity of single layer graphene domains are significantly increased and decreased, respectively, compared with the sample prepared by the "_face-to-face_" method. Graphene grown with both methods show the same lattice constant as extracted from the distance between LEED spots, indicating that the different growth methods do not affect the atomic structure of graphene.

Direct measurements of the electronic band structure were performed by ARPES as shown in Fig. 6. The characteristic conical energy versus momentum dispersion is observed [34]. As previously reported [35, 36] interference effects between photo-emitted electrons hinder the observation of both sides of the conical bands along the KM direction (see horizontal line in the Brillouin zone the cartoon). Instead, a single linear band ($\pi$ band) is observed along this direction in single layer graphene grown at 1530 °C as shown in panel (a), in good agreement with Ref. 37 [38]. The $\pi$ band splits into two and three bands for double layer (grown at 1600 °C (panel (b)) and triple layer (grown at 1700 °C panel (c)), respectively, due to interlayer coupling. For a quantitative evaluation, momentum and energy distribution curves (MDC) and (EDC) for each graphene sample are analyzed. It is found that Dirac point shifted below the Fermi level by 0.36 eV for single layer, 0.27 eV for bilayer and 0.21 eV for trilayer, which is consistent with previous reports [39].

Raman spectroscopy is a simple and efficient way to characterize quality and identify thickness of graphene in few layers. Fig. 7 (a) and (b) present the spectra around the D and G peak of "*face-to-face*" grown and UHV-grown single layer epitaxial graphene on 6H-SiC (0001). For both graphene samples, the D and G peaks are observed at 1362 cm$^{-1}$ and 1590 cm$^{-1}$, respectively. Comparing to UHV-grown graphene, the defect D peak of monolayer sample prepared by this new method is much weaker, attesting to the better quality. Fig. 7 (c) displays Raman spectra of the 2D peaks of mono-, bi- and trilayer graphene grown by "*face-to-face*" method, respectively. As the thickness of graphene increases from 1 to 3 layers, a single Lorentzian peak centered at 2709 cm$^{-1}$ shifts to lower wavenumber 2690 cm$^{-1}$, Meanwhile, the peak widths FWHM (2D) rise from 35 cm$^{-1}$ to 64 cm$^{-1}$ and 75 cm$^{-1}$. As shown in the figure, monolayer epitaxial graphene can be fitted with single Lorentzian (green line). However, in case of bilayer graphene, decomposition into four Lorentzians is necessary, which is well known as a characteristic property for indentifying two layer graphene. As thickness increases to 3, 2D peak becomes much broader and contains multiple components. Above Raman result is consistent with previous reports [40, 41] and the ARPES results here shown.

In summary, we developed a new method for synthesizing micrometer scale graphene sheets by simultaneous vacuum thermal decomposition of two SiC substrates directly facing each other with only a narrow 25 micron gap between them. The unique sample geometry provides a strikingly simple method for restricting the effective Si sublimation rate enabling precise control of graphene growth. By manipulating annealing temperature, the thickness of the graphene is controlled, which is confirmed by ARPES and Raman spectroscopy. AFM measurements demonstrate that homogeneity of graphene is notably improved by the "*face-to-face*" method compared with the conventional method. We stress that the method is straightforward and does not require sophisticated fabrication, or elaborate specifications or additional materials for the restriction of the Si sublimation rate. Thus this method may have significant potential for the efficient production of wafer size graphene, an essential component to develop graphene based devices.

We would like to thank D. A. Siegel for useful discussions, B. S. Geng and F. Wang for help with the Raman measurement. This work was supported by the Director, Office of Science, Office of Basic Energy Sciences, Materials Sciences and Engineering Division of the U.S. Department of Energy under Contract No. DE-AC02_05CH11231. X. Z. Yu would like to thank her advisor, Prof. Wenzhong Shen in Shanghai Jiao Tong University for providing her the opportunity to spend a period in UC Berkeley. Such stay was supported by the joint-training project between Berkeley and China by the China Scholarship Council.

# FIGURE CAPTIONS

Fig.1: LEED patterns with a primary energy of 180 eV, obtained at four different stages during the growth of sample A. (a) 1 × 1 spots of SiC, after a 5 min anneal around 1000 ℃ followed by the initial cleaning procedure under Si fl a. (b) (√3 × √3)R30 reconstruction, after 5 min around 1100 ° C. (c) (6√3 × 6√3)R30 reconstruction, after 10 min around 1250 ° C.

Fig. 2: LEEM image of graphene on 6H-SiC(0001) with a nominal thickness of 1.2 ML formed by annealing in Ar (P = 900 mbar, T = 1650 ℃). Copyright 2009 Nature.

Fig.3: (a) Schematic view of the configuration used for "*fact-to-face*" growth method setup; (b) magnified view of the sample setup highlighted in panel (a), where rectangulars in red represent SiC substrates, and lines in blue represent Ta foil; (c) magnified view of mounted SiC substrates highlighted by red lines in panel (b), the distance between Si-face of the two substrates equals to the thickness of Ta foil (d = 25μm). SiC substrate is in light gray and Ta foil is in dark gray.

Fig. 4: Schematic view of the two-step growth model for preparing epitaxial graphene

Fig. 5: Atomic force microscopy images of the topographies of (a) pristine SiC surface after annealing at 700℃ for 4 hours, (b) SiC surface following heating at 1500℃, (c) single layer graphene grown on SiC surface after annealing at 1530℃ for 20 minutes, and (d) single layer graphene prepared in the same way as described in Ref.18 [18]. The insets in panels (b)-(d) are corresponding LEED patterns measured at room temperature with a primary energy of 98.9eV.

Fig. 6: Band structure of single layer, double layer, and triple layer graphene epitaxially grown on 6H-SiC (0001) obtained by ARPES using He II radiation. Inset of (a) shows the Brillouin zone of graphene. The horizontal line along the KM direction shows the measurement geometry in *k*-space. The position of Dirac energy is indicated by yellow arrow. Momentum distribution curves (MDC) extracted from the images at the energies indicated by the red lines (-0.8 eV).

Fig. 7: (a - b) Comparison of Raman spectra around the D and G peaks of "*fact-to-face*" grown and UHV-grown epitaxial graphene on 6H-SiC (0001). (c) Raman spectra of the 2D peaks of monolayer, bilayer and trilayer graphene, respectively. The 2D peak of monolayer graphene spectrum can be fitted by one Lorentzian line (red line), the 2D peak of bilayer graphene can be deconvoluted in four components (green dash lines).

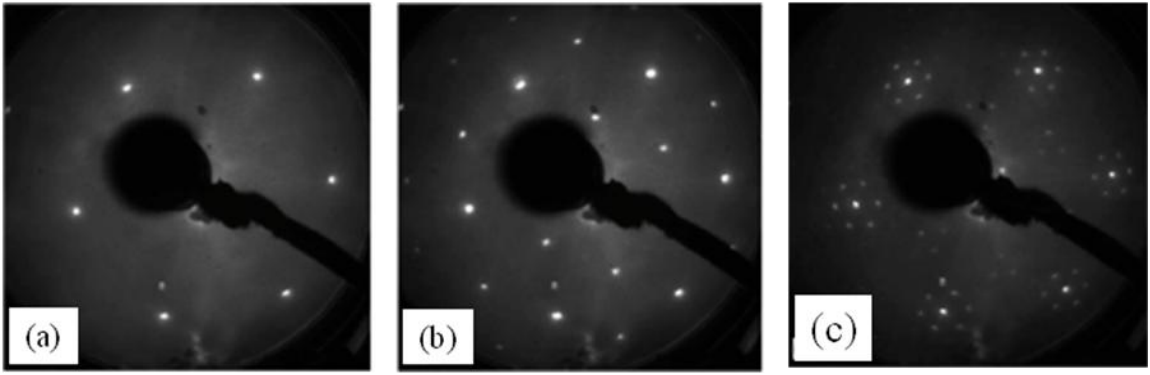

Fig.1 (a–c) LEED patterns with a primary energy of 180 eV, obtained at four different stages during the growth of sample A. (a) 1 × 1 spots of SiC, after a 5 min anneal around 1000 °C followed by the initial cleaning procedure under Si flux. (b) (√3 × √3)R30 reconstruction, after 5 min around 1100 ° C. (c) (6√3 × 6√3)R30 reconstruction, after 10 min around 1250 ° C.

X. Z. Yu, *et al.*, Fig. 1 of 7

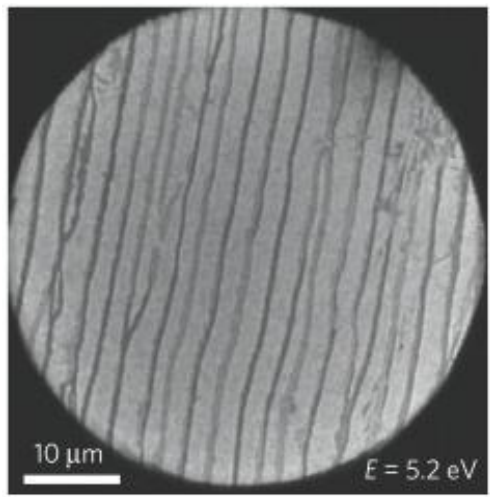

Fig.2 LEEM image of graphene on 6H-SiC(0001) with a nominal thickness of 1.2 ML formed by annealing in Ar (P=900 mbar, T=1650℃). Copyright 2009 Nature.

X. Z. Yu, *et al.*, Fig. 2 of 7

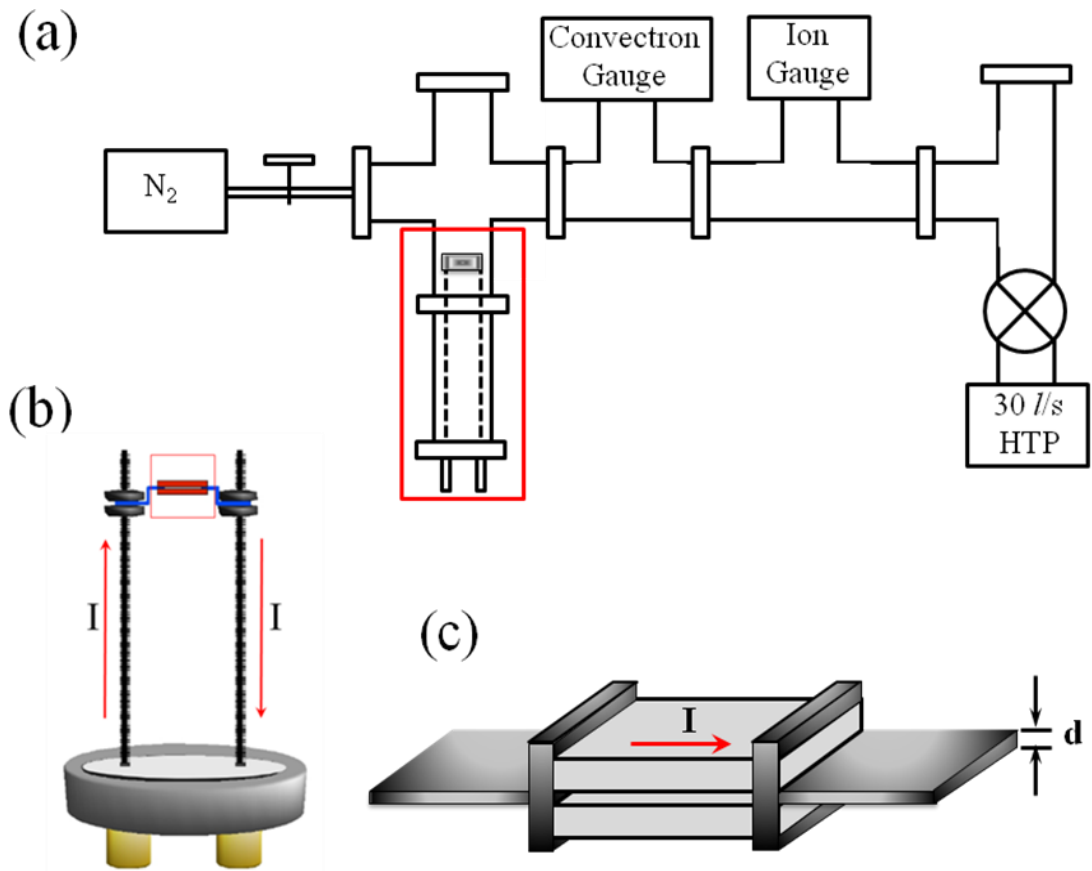

Fig.3 (a) Schematic view of the configuration used for "*face-to-face*" growth method setup; (b) magnified view of the sample setup highlighted in panel (a), where rectangulars in red represent SiC substrates, and lines in blue represent Ta foil; (c) magnified view of mounted SiC substrates highlighted by red lines in panel (b), the distance between Si-face of the two substrates equals to the thickness of Ta foil (d = 25μm). SiC substrate is in light gray and Ta foil is in dark gray.

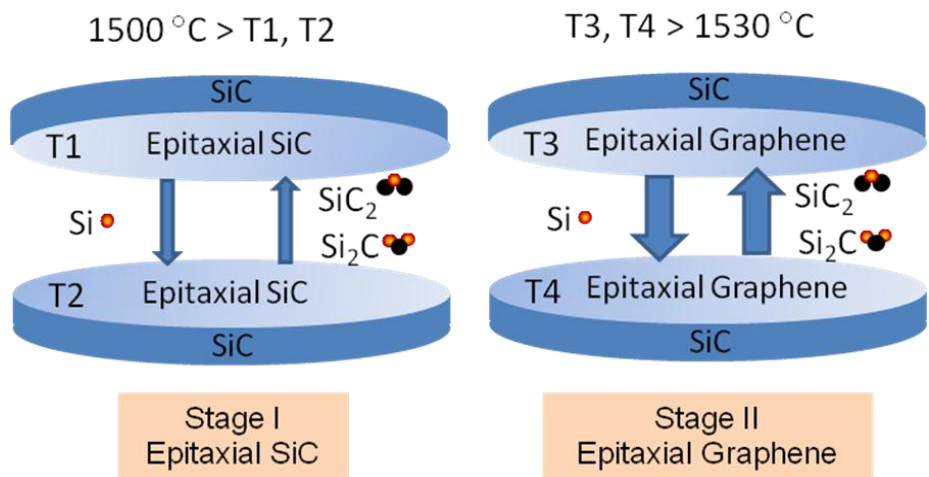

Fig. 4: Schematic view of the two-step growth model for preparing epitaxial graphene

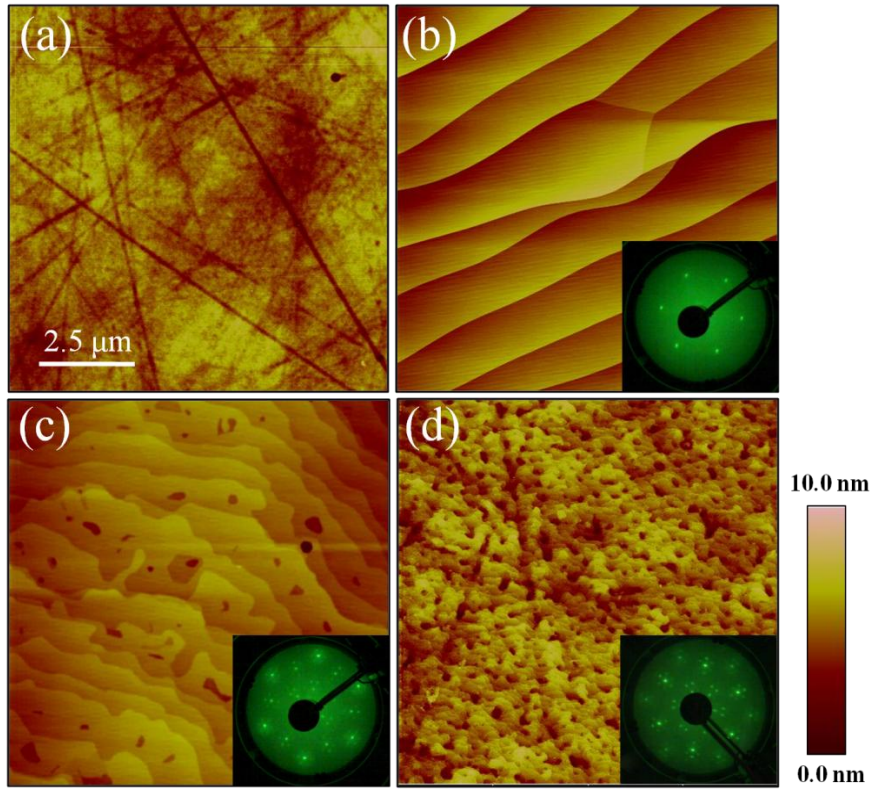

Fig. 5: Atomic force microscopy images of the topographies of (a) pristine SiC surface after annealing at 700°C for 4 hours, (b) SiC surface following heating at 1500°C, (c) single layer graphene grown on SiC surface after annealing at 1530°C for 20 minutes, and (d) single layer graphene prepared in the same way as described in Ref. 16 [16]. The insets in panels (b)-(d) are corresponding LEED patterns measured at room temperature with a primary energy of 98.9eV

X. Z. Yu, *et al.*, Fig. 5 of 7

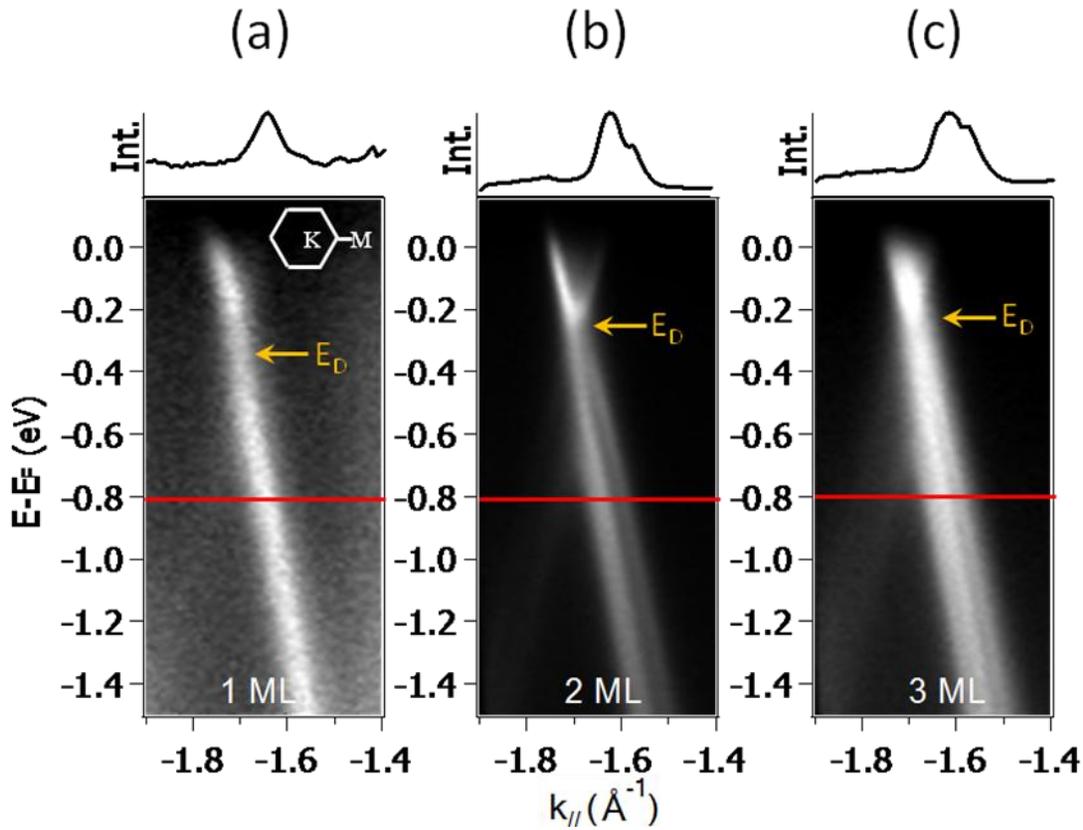

Fig. 6: Band structure of single layer, double layer, and triple layer graphene epitaxially grown on 6H-SiC (0001) obtained by ARPES using He II radiation. Inset of (a) shows the Brillouin zone of graphene. The horizontal line along the KM direction shows the measurement geometry in *k*-space. The position of Dirac energy is indicated by yellow arrow. Momentum distribution curves (MDC) extracted from the images at the energies indicated by the red lines (-0.8 eV).

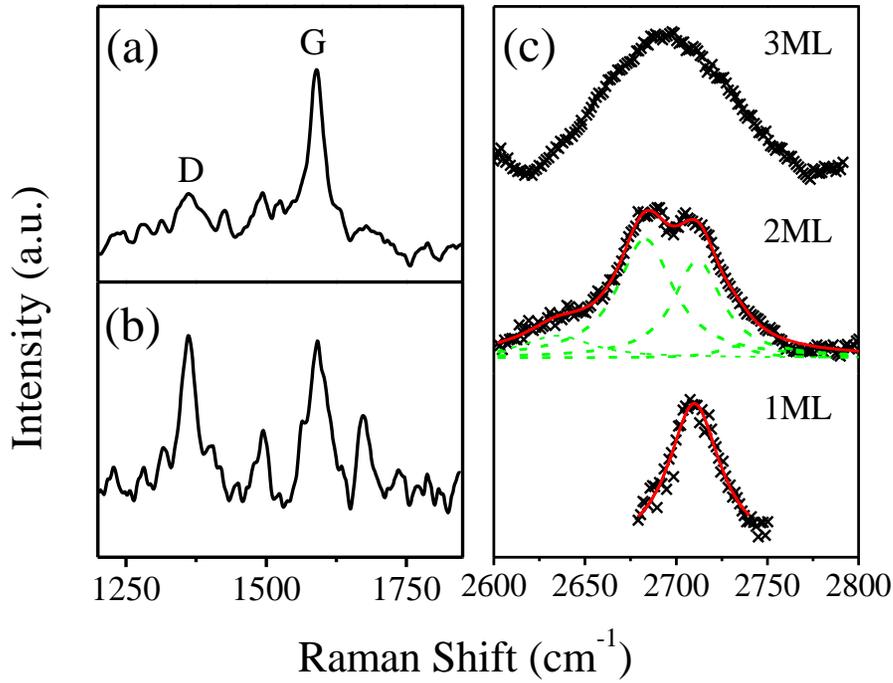

Fig. 7: (a-b) Comparison of Raman spectra around the D and G peaks of "*face-to-face*" grown and UHV-grown epitaxial graphene on 6H-SiC (0001). (c) Raman spectra of the 2D peaks of monolayer, bilayer and trilayer graphene, respectively. The 2D peak of monolayer graphene spectrum can be fitted by one Lorentzian line (red line), the 2D peak of bilayer graphene can be deconvoluted in four components (green dash lines).

X. Z. Yu, *et al.*, Fig. 7 of 7